\documentclass[
preprintnumbers, 
showpacs, 
amsmath, 
showkeys, 
amssymb, 
aps, 
superscriptaddress, 
twocolumn, 
longbibliography, 
letterpaper,
]{revtex4-2}
\usepackage{lipsum, babel}
\usepackage{color}
\usepackage{hyperref}
\usepackage{bm}
\usepackage{amsfonts}
\usepackage{mathrsfs}
\usepackage{graphicx}
\usepackage{amsmath}
\usepackage{float}
\usepackage{braket}
\usepackage{natbib}
\usepackage{orcidlink}

     \hypersetup{
    colorlinks=true,
    linkcolor=red,
    citecolor=blue,
}

\begin{document}

\newcommand{\sgn}{\operatorname{sgn}}
\newcommand{\hhat}[1]{\hat {\hat{#1}}}
\newcommand{\pslash}[1]{#1\llap{\sl/}}
\newcommand{\kslash}[1]{\rlap{\sl/}#1}
\newcommand{\lab}[1]{}
% to create labels.
%\newcommand{\iref}[2]{\footnote{\hyperlink{lb:#1}{\textit{$^\spadesuit$#2}}}}
%\newcommand{\iref}[2]{}
% reference to other part in this notes.
\newcommand{\sto}[1]{\begin{center} \textit{#1} \end{center}}
\newcommand{\rf}[1]{{\color{blue}[\textit{#1}]}}
% Reference
\newcommand{\eml}[1]{#1}
% Emphasis
\newcommand{\el}[1]{\label{#1}}
% Equation labeling
\newcommand{\er}[1]{Eq.\eqref{#1}}
% Equation Reference
\newcommand{\df}[1]{\textbf{#1}}
% Temporarily replace \textbf
\newcommand{\mdf}[1]{\pmb{#1}}
% Use for vectors etc.
\newcommand{\ft}[1]{\footnote{#1}}
% footnote.
\newcommand{\n}[1]{$#1$}
% Use for numbers etc.
% \newcommand{\cjktext}[1]{\begin{CJK}{GB}{gbsn} #1 \end{CJK}} 
% Language support
\newcommand{\fals}[1]{$^\times$ #1}
% wrong statement
\newcommand{\new}{{\color{red}$^{NEW}$ }}
% update
% \newcommand{\ci}[1]{\cite{#1}}
\newcommand{\ci}[1]{}
\newcommand{\de}[1]{{\color{green}\underline{#1}}}
\newcommand{\ke}{\rangle}
\newcommand{\br}{\langle}
\newcommand{\lb}{\left(}
\newcommand{\rb}{\right)}
\newcommand{\lbk}{\left[}
\newcommand{\rbk}{\right]}
\newcommand{\blb}{\Big(}
\newcommand{\brb}{\Big)}
\newcommand{\nn}{\nonumber \\}
\newcommand{\p}{\partial}
\newcommand{\pd}[1]{\frac {\partial} {\partial #1}}
\newcommand{\cd}{\nabla}
\newcommand{\cc}{$>$}
% ##### ###### ###### ######
\newcommand{\bqa}{\begin{eqnarray}}
\newcommand{\eqa}{\end{eqnarray}}
\newcommand{\bqe}{\begin{equation}}
\newcommand{\eqe}{\end{equation}}
\newcommand{\bay}[1]{\left(\begin{array}{#1}}
\newcommand{\eay}{\end{array}\right)}
\newcommand{\eg}{\textit{e.g.} }
\newcommand{\ie}{\textit{i.e.}, }
\newcommand{\iv}[1]{{#1}^{-1}}
\newcommand{\st}[1]{|#1\ke}
\newcommand{\at}[1]{{\Big|}_{#1}}
\newcommand{\zt}[1]{\texttt{#1}}
\newcommand{\non}{\nonumber}
\newcommand{\m}{\mu}
% ##### ###### ###### ######
% Greek Letters
\def\xa{{m}}
\def\xA{{m}}
\def\xb{{\beta}}
\def\xB{{\Beta}}
\def\xd{{\delta}}
\def\xD{{\Delta}}
\def\xe{{\epsilon}}
\def\xE{{\Epsilon}}
\def\xve{{\varepsilon}}
\def\xg{{\gamma}}
\def\xG{{\Gamma}}
\def\xk{{\kappa}}
\def\xK{{\Kappa}}
\def\xl{{\lambda}}
\def\xL{{\Lambda}}
\def\xo{{\omega}}
\def\xO{{\Omega}}
\def\xvp{{\varphi}}
\def\xs{{\sigma}}
\def\xS{{\Sigma}}
\def\xt{{\theta}}
\def\xvt{{\vartheta}}
\def\xT{{\Theta}}
% ##### ###### ###### ######
\def \Tr {{\rm Tr}}
\def\CA{{\cal A}}
\def\CC{{\cal C}}
\def\CD{{\cal D}}
\def\CE{{\cal E}}
\def\CF{{\cal F}}
\def\CH{{\cal H}}
\def\CJ{{\cal J}}
\def\CK{{\cal K}}
\def\CL{{\cal L}}
\def\CM{{\cal M}}
\def\CN{{\cal N}}
\def\CO{{\cal O}}
\def\CP{{\cal P}}
\def\CQ{{\cal Q}}
\def\CR{{\cal R}}
\def\CS{{\cal S}}
\def\CT{{\cal T}}
\def\CV{{\cal V}}
\def\CW{{\cal W}}
\def\CY{{\cal Y}}
\def\BC{\mathbb{C}}
\def\BR{\mathbb{R}}
\def\BZ{\mathbb{Z}}
\def\sA{\mathscr{A}}
\def\sB{\mathscr{B}}
\def\sF{\mathscr{F}}
\def\sG{\mathscr{G}}
\def\sH{\mathscr{H}}
\def\sJ{\mathscr{J}}
\def\sL{\mathscr{L}}
\def\sM{\mathscr{M}}
\def\sN{\mathscr{N}}
\def\sO{\mathscr{O}}
\def\sP{\mathscr{P}}
\def\sR{\mathscr{R}}
\def\sQ{\mathscr{Q}}
\def\sS{\mathscr{S}}
\def\sX{\mathscr{X}}

% 你的 ORCID 链接

% 使用命令插入 ORCID 链接

\author{Wei Kou\orcidlink{0000-0002-4152-2150}}
\email{kouwei@impcas.ac.cn}

\affiliation{Institute of Modern Physics, Chinese Academy of Sciences, Lanzhou 730000, China}
\affiliation{School of Nuclear Science and Technology, University of Chinese Academy of Sciences, Beijing 100049, China}
\author{Xurong Chen}
\email{xchen@impcas.ac.cn (Corresponding Author)}
\affiliation{Institute of Modern Physics, Chinese Academy of Sciences, Lanzhou 730000, China}
\affiliation{School of Nuclear Science and Technology, University of Chinese Academy of Sciences, Beijing 100049, China}
\affiliation{Southern Center for Nuclear Science Theory (SCNT), Institute of Modern Physics, Chinese Academy of Sciences, Huizhou 516000, Guangdong Province, China}

\title{Machine Learning Insights into Quark-Antiquark Interactions: Probing Field Distributions and String Tension in QCD}
%\date{\today}

\begin{abstract}
Understanding the interactions between quark-antiquark pairs is essential for elucidating quark confinement within the framework of quantum chromodynamics (QCD). This study investigates the field distribution patterns that arise between these pairs by employing advanced machine learning techniques, namely multilayer perceptrons (MLP) and Kolmogorov-Arnold networks (KAN), to analyze data obtained from lattice QCD simulations. The models developed through this training are then applied to calculate the string tension and width associated with chromo flux tubes, and these results are rigorously compared to those derived from lattice QCD. Moreover, we introduce a preliminary analytical expression that characterizes the field distribution as a function of quark separation, utilizing the KAN methodology. Our comprehensive quantitative analysis underscores the potential of integrating machine learning approaches into conventional QCD research.
\end{abstract}

%\pacs{04.60.Bc, 04.62.+v, 04.70.Dy}

\maketitle

\section{Introduction}
\label{sec:introduction}
The development of QCD has been pivotal, offering key examples for hadronic physics theory and phenomenology. Despite significant progress, explaining quark confinement remains a challenge in strong interaction studies \cite{greensite2011introduction}. Lattice QCD \cite{Gupta:1997nd}, using Monte Carlo simulations on four-dimensional spacetime lattices, tackles many non-perturbative QCD problems with advanced computational tools, bolstering confidence in these methods. Non-Abelian gauge fields exhibit flux tube structures between quark-antiquark pairs due to gluon self-interactions. This framework, incorporating string model elements, supports numerical and model analyses of strong interactions.

Extensive foundational studies on flux tubes in QCD have established a robust framework, encompassing theoretical analyses and lattice QCD simulations \cite{DiGiacomo:1989yp,Fukugita:1983du,Flower:1985gs,Cea:1992sd,Cea:1995zt,Bali:1994de,Luscher:1980iy,Baker:1989qp,Haymaker:1994fm,Cea:2012qw,Baker:2019gsi,Galsandorj:2023rqw,Baker:2024peg,Baker:2024rjq}. Investigations into how flux tube profiles depend on the separation of quark-antiquark pairs have illuminated properties such as confinement, deconfinement, and QCD string breaking \cite{Philipsen:1998de,Kratochvila:2002vm,Bali:2005fu}. Lattice simulations provide visualizations of flux tube profiles between static quark-antiquark pairs and detailed distributions of chromoelectric and chromomagnetic fields. The transverse chromoelectric field profile at the midpoint of the quark-antiquark pair is particularly noteworthy. By examining field distributions at varying separations, distance-dependent characteristics can be identified. The chromofield generated by static $q\bar{q}$ pairs is quantified using correlation functions. For theoretical and lattice setup details, see Ref. \cite{Baker:2019gsi}.

Monte Carlo simulations visualize the chromo field $E(d,x_t)$ across the transverse plane coordinate $x_t$ for a given quark separation $d$. While the data points are discrete, they outline an approximate distribution. A parametric model can link the chromo flux tube structure to the dual Meissner effect in the dual superconducting confinement model \cite{Clem:1975jr,Cea:1992sd}. Detailed discussions on the chromo field are provided later in this paper. A unified description of the chromo field using a general 2D function $(d, x_t)$ remains unresolved. Current methods fit parameters for each quark separation $d$, yielding distinct parameter sets for different $d$, which is convenient for deriving continuous chromo field functions, as shown in \cite{Kharzeev:2014xta,Iritani:2015zwa,Kou:2024nca,Kou:2024dml}. However, due to computational limits, lattice simulations cannot cover all possible separations, especially at large distances, where the signal-to-noise ratio becomes excessively high \cite{Baker:2024peg}. These challenges emphasize the need for a novel approach to describe chromo field distributions across various quark separations.

Advances in artificial intelligence (AI), particularly in machine learning and deep learning, have benefited numerous fields in recent years, including physics. Neural networks have become a well-established tool for analyzing high-energy physics data \cite{Forte:2002fg}, while deep learning is increasingly applied in high-energy phenomenology \cite{Larkoski:2017jix,Guest:2018yhq,Radovic:2018dip,Albertsson:2018maf,Carleo:2019ptp,Bourilkov:2019yoi,Schwartz:2021ftp,Karagiorgi:2021ngt,Boehnlein:2021eym,Shanahan:2022ifi,Yang:2022yfr,Li:2022ozl,He:2023zin,Zhou:2023pti,Zhou:2023tvv,Pang:2024kid,Ma:2023zfj,Luo:2024iwf,Chen:2024epd,OmanaKuttan:2023bnb,Shi:2022vfr,Shi:2022fei,Shi:2021qri,Mansouri:2024uwc,Chen:2024mmd,Chen:2024ckb,Wang:2023poi}. Deep learning enhances flexibility in modeling data not easily described by explicit functions, even though its mechanisms often resemble a “black box” and remain difficult to fully interpret. Traditionally, model development for explaining experimental data or lattice QCD results has been theory-driven. In contrast, machine learning methods allow for the deduction of physical model features from data, constituting an “inverse problem” strategy. This data-driven approach reduces the risk of overly strong assumptions in model construction but still relies on human guidance for capturing physical characteristics. Neural networks based on multi-layer perceptrons (MLP) are commonly employed for parameter estimation, akin to global data fitting. Recently, the Kolmogorov-Arnold networks (KAN) method was introduced to the open-source community \cite{liu2024kankolmogorovarnoldnetworks,liu2024kan20kolmogorovarnoldnetworks}, offering advantages in interpretability over MLP for small-scale scientific tasks and allowing human intervention during training. This study examines the similarities and differences between these two neural network methods in describing the flux tube profile between quark-antiquark pairs, with particular attention to how analytical expressions capture the distance dependence of the chromo field distribution. In particular, the latter aspect is currently difficult to achieve with other parameterization schemes, which can only fit a set of parameters for specific quark-antiquark separations, resulting in poor continuity. Through deep learning, researchers can derive more continuous model representations from limited information or data. The paper is organized as follows: Sec. \ref{sec:Form} reviews the chromo field distribution and introduces relevant aspects of machine learning. Sec. \ref{sec:result} presents the main machine learning results and numerical computations. The conclusion discusses findings and future directions.

\section{Chromoelectric Field and Machine Learning}
\label{sec:Form}
\subsection{String tension of chromoelectric field}
\label{sec:2.1}
The description of (3+1)-dimensional non-Abelian gauge fields is complex. To simplify the framework, a quasi-Abelian picture based on the dual Meissner effect has been proposed, which provides support for numerous lattice QCD studies \cite{Singh:1993jj,Schilling:1998gz,Chernodub:2000rg,Chernodub:2005gz,Suzuki:2009xy,Cea:2012qw}. This is also referred to as the``maximum Abelian projection." These arguments are summarized and reviewed in Ref. \cite{Kharzeev:2014xta}, where the relationship between confinement and vacuum magnetic monopole condensation is discussed. In a simple string model, confined quarks are connected by Abelian chromoelectric flux tubes. These flux tubes are dual to Abrikosov-Nielsen-Olesen (ANO) vortices \cite{Cea:2012qw,Abrikosov:1956sx,Nielsen:1973cs} in type-II superconductors, and their dynamical behavior can be described by (1+1)-dimensional effective theory \cite{Witten:1984eb}. 

In lattice simulations, the distribution behavior of the 3+1-dimensional chromoelectric flux tube is captured. In simple terms, the flux tube field strength $E(x_t)$ between quark and antiquark pairs varies with the transverse coordinates, and the tension of the flux tube $\sigma_T$ can be described by the total contribution of the chromoelectric field \cite{Kharzeev:2014xta,Kou:2024nca,Kou:2024dml}
\begin{equation}
	\sigma_T\simeq\frac12\int d^2x_t E^2(x_t).
	\label{eq:tension}
\end{equation}
Clearly, the information on the (3+1)-dimensional flux tubes can be obtained through lattice simulations, similar to previous works \cite{Kou:2024nca,Kou:2024dml}, where the extracted flux tube data from different static color sources in Ref. \cite{Baker:2019gsi} are utilized. The measured chromoelectric field as a function of the transverse coordinate can be well described by the following parameterization \cite{Clem:1975jr}
\begin{equation}
	E(x_t)=\frac\varphi{2\pi}\frac{\mu^2}\alpha\frac{K_0[(\mu^2x_t^2+\alpha^2)^{1/2}]}{K_1[\alpha]},
	\label{eq:chromofileld}
\end{equation}
where $\varphi$, $\alpha$, and $ \mu $ are fitting parameters. Moreover, $K_n(x) $ is the modified Bessel function of the second kind. The parameter $\varphi $ represents the external flux, while $\mu $ is the inverse of the London penetration depth $ \lambda $.  Equation (\ref{eq:chromofileld}) is a simple analytical expression derived from Ampère’s law and the Ginzburg-Landau equation. In particular, it can be simplified to the London model outside the vortex core in the presence of a transverse magnetic field distribution. This expression reflects the compatibility of color flux tubes with dual superconductor theories.

In previous outstanding works, researchers often utilize Eq. (\ref{eq:chromofileld}) to fit lattice simulation data, obtaining different parameter sets for varying quark-antiquark separation distances $d$. Analyzing the optimized parameters derived from the fits allows for a rough understanding of how the color field distribution changes with $d$. This work also considers the results from lattice simulations \cite{Baker:2019gsi}, which include samples of color field distributions for physical separations between quark and antiquark pairs ranging approximately from 0.37 fm to 1.2 fm. It is important to emphasize that the purpose of this work is to introduce neural networks for the analysis of these data. To facilitate comparisons with widely accepted results, we select only the non-perturbative portion of the flux tube data, also referred to as the $curl\ procedure$ \cite{Baker:2019gsi}, which is obtained by subtracting the contributions of the effective Coulomb field from the total gauge field contributions.

\subsection{MLP vs KAN}
\label{sec:2.2}
\begin{figure}[htbp]
	\begin{center}
		\includegraphics[width=0.96\linewidth]{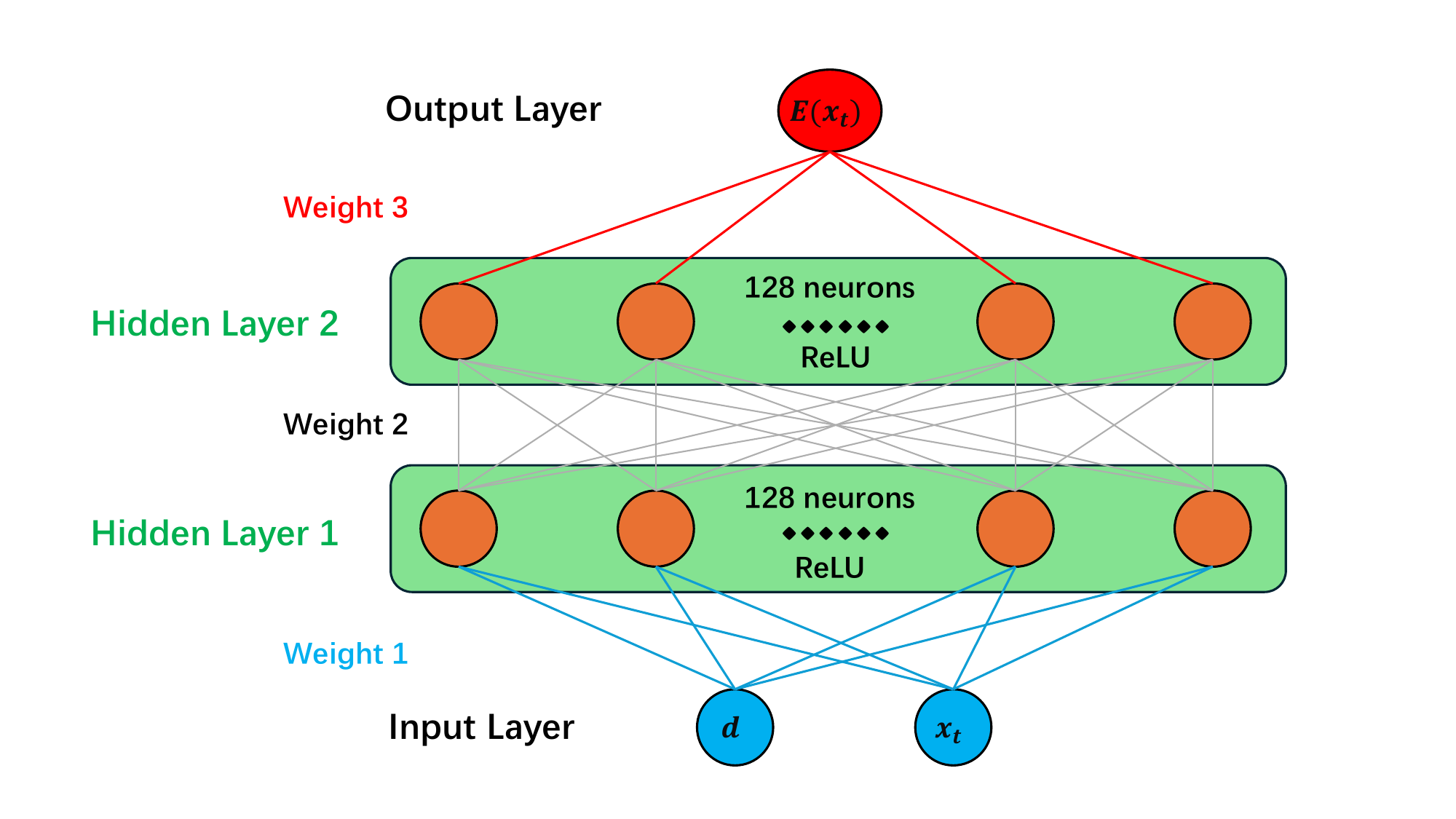}
	\end{center}
	\caption{The schematic representation of training the lattice simulation data using the MLP architecture. Each of the two hidden layers contains 128 neurons, with a complete mapping between the layers. The input layer consists of two neurons that encode information about the quark-antiquark pair separation distance $d$ and the transverse coordinate $x_t$, while the output layer corresponds to the desired color field strength $E(x_t)$.}
	\label{fig:MLP}
\end{figure}

The MLP \cite{hornik1989multilayer,cybenko1989approximation,haykin1998neural} serves as the most fundamental constituent of deep learning to date and stands as the default model in machine learning for approximating nonlinear functions, making it particularly significant. However, its relatively low interpretability presents some drawbacks. As a benchmark, we utilize the advantages of the MLP to conduct the core aspects of this work. The MLP architecture employed in this study includes an input layer, two hidden layers, and an output layer. Given that our training dataset, derived from lattice simulations of the color field distribution, depends on two variables $d$ and $x_t$, the input layer requires two neurons, while the output layer, representing $E(x_t)$, necessitates a single neuron. A schematic representation of the specific architecture is depicted in Fig. \ref{fig:MLP}. 

In the MLP architecture of this work, the number of hidden layers is set to 2, each containing 128 neurons, and the activation function used is ReLU. This architecture is relatively straightforward to understand, and hyperparameters such as regularization, loss function, and learning rate can be adjusted according to the specific requirements of the training dataset. In this work, we configure the Mean Square Error (MSE) between the model predictions and lattice simulation results as the loss function, incorporating an L2 regularization scheme. To further validate the reliability of the training process and avoid overfitting,we randomly select $20\%$ of the data from the entire dataset as a test set for each training session. The trained model is required to describe both the test and training sets.

Another architecture, KAN, relies on the Kolmogorov-Arnold Representation theorem (KART), also known as the superposition theorem. The work of Kolmogorov and Arnold establishes that a multivariate continuous function can be expressed as a finite combination of binary sums of continuous univariate functions \cite{liu2024kankolmogorovarnoldnetworks}. Compared to the MLP framework, KAN presents advantages. This is attributable to the fact that the activation functions in the MLP architecture are fixed at the nodes, while the weight functions are learnable. In contrast, the KAN architecture features learnable activation functions at the weight edges. Additionally, KAN completely eliminates linear weight functions, replacing them with parameterized spline univariate functions \cite{liu2024kankolmogorovarnoldnetworks}. The KAN architecture for deep learning is still in the early stages of development; however, it continues to attract significant attention in the field of computer science research. A detailed comparison of the strengths and weaknesses between KAN and traditional MLP is ongoing, and our aim is to evaluate their performance on the same problem through the specific physical issue of chromo flux tube distribution in QCD.

\begin{figure}[htbp]
	\begin{center}
		\includegraphics[width=0.98\linewidth]{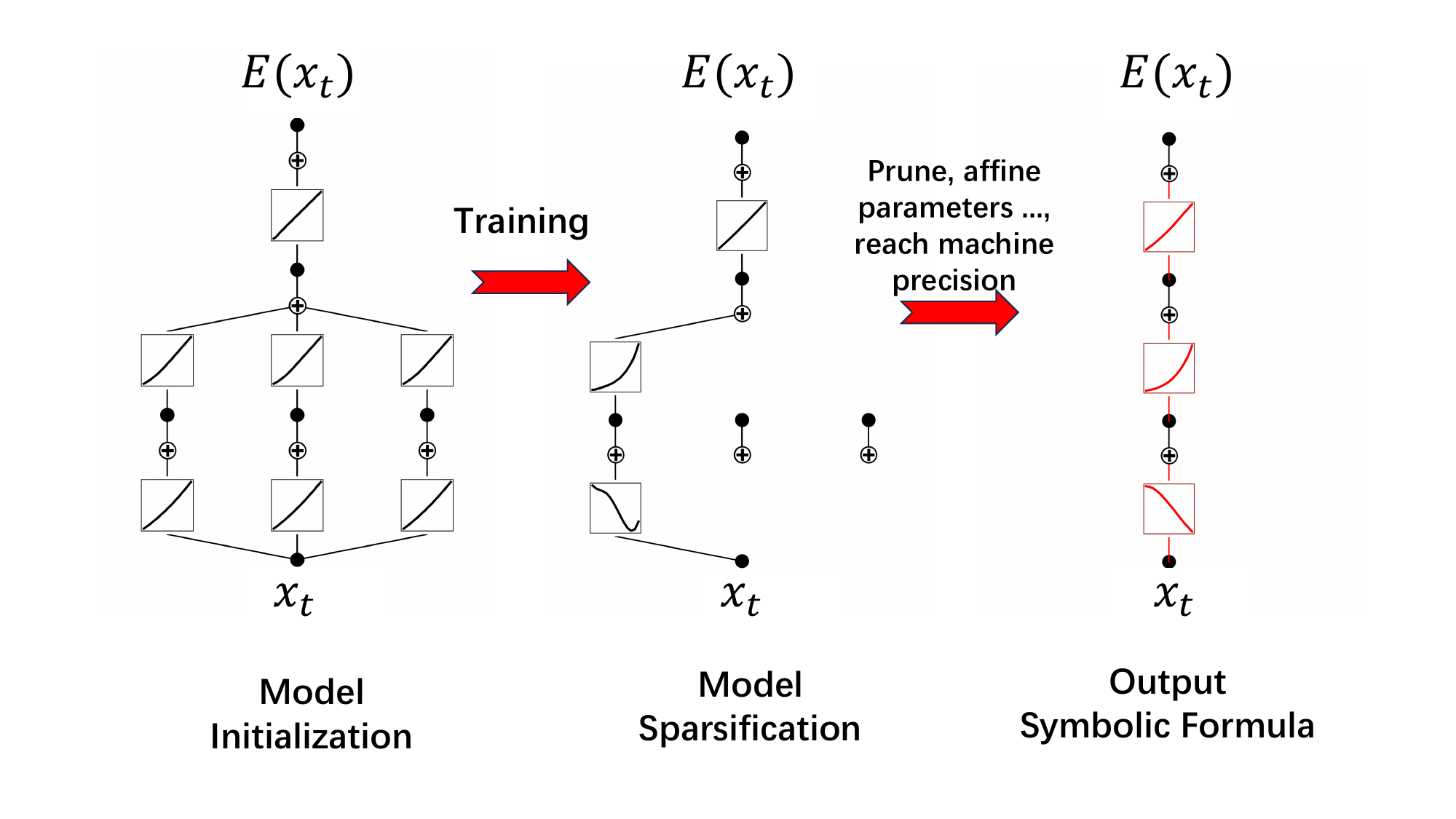}
	\end{center}
	\caption{The schematic representation of the KAN architecture training on a univariate dataset is shown. The red curve represents the determined function combination.}
	\label{fig:KAN}
\end{figure}

Using a univariate function dataset as an example, the KAN architecture is relatively easy to understand. We take the chromo field distribution for a quark-antiquark separation distance of $d = 0.51$ fm, as reported in \cite{Baker:2019gsi}, as our training set. The input layer consists of a single neuron representing the $x_t$ tensor network. We employ the KAN architecture configured as $(1,3,1,1)$ for mapping, as illustrated in Fig. \ref{fig:KAN}. The first step involves initializing the spline functions, which is essential for establishing a robust training process. Subsequently, we introduce an optimization algorithm to continue training and identify the features across the entire dataset. This process can be conducted in multiple steps, allowing for adjustments to the training strategy at each stage, ultimately achieving machine precision. Finally, we generate a symbolic expression for the target function based on the network of the final model, a step that corresponds to linear regression fitting. A feasibility analysis of training unit functions using this architecture will be provided in the next section.

\section{Results and discussions}
\label{sec:result}
This section is divided into two parts. First, we conduct a feasibility test of the two machine learning architectures, MLP and KAN, as mentioned earlier, using the dataset of chromo field distribution for $d = 0.51$ fm from \cite{Baker:2019gsi}. This distribution can be effectively parameterized as shown in Eq. (\ref{eq:chromofileld}), we also include a comparison of the parameterization results. Secondly, we modify the input layer to consist of two elements, $d$ and $x_t$, incorporating all chromo fields corresponding to various $d$ values from \cite{Baker:2019gsi} into the training set, thereby constructing an overall tensor network to apply MLP and KAN for separate training. Notably, we assume that the chromo field distribution is independent of the longitudinal coordinate; however, in practice, the variations in the longitudinal coordinate have a negligible impact on the chromo field distribution \cite{Baker:2019gsi,Cea:2012qw}. In addition to the comparison of chromo field distributions, we also provide a comparison of the calculations of flux tube string tension and the root mean square width of the flux tubes.

\subsection{Feasibility assessment of the two machine learning architectures}
\label{subsec: 1d}
The implementation of the MLP framework is illustrated in Fig. \ref{fig:MLP} (note that the input layer consists of a single neuron), where a standard training process can be readily found within the machine learning community. In contrast, the KAN architecture includes an additional step, specifically the determination of a symbolic representation of the objective function. For testing, the neuron structure of KAN is configured as $(1,3,3,1)$, as shown in Fig. \ref{fig:KAN}. After initialization, each node can map to a spline function, which is the learnable component. Following one round of training, it is evident that only a single function path remains, a result of the penalty mechanism employed to eliminate high-frequency oscillations and highly irregular paths. Once a unique function path is obtained, pruning and refinement training can be performed, and upon reaching machine precision, linear regression can be conducted to determine the target symbolic function and its corresponding parameters. It is noteworthy that the loss functions corresponding to both the MLP and KAN frameworks in the univariate function training tests converge to  $\sim 10^{-5}$.

\begin{figure}[htbp]
	\begin{center}
		\includegraphics[width=0.98\linewidth]{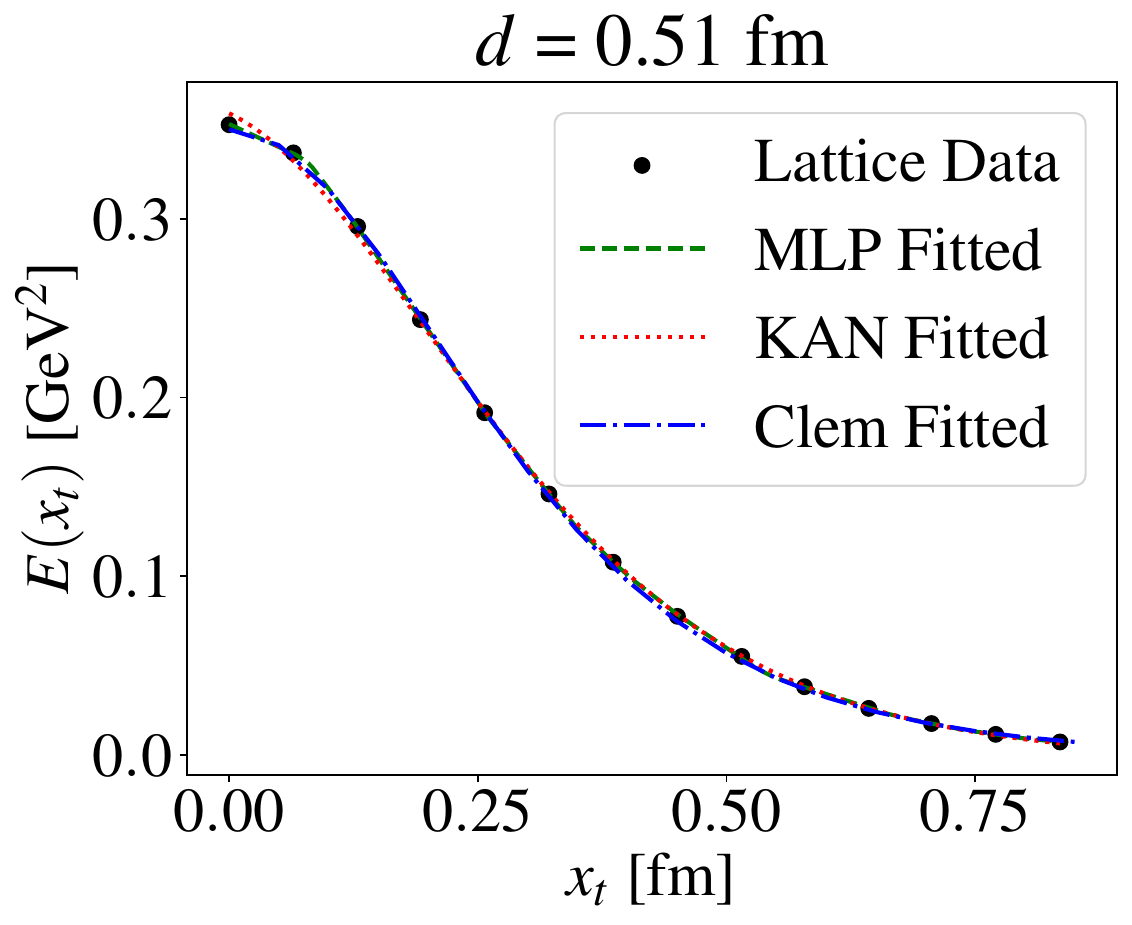}
	\end{center}
	\caption{The transverse distribution of the chromo field for quark-antiquark separation at $d = 0.51$ fm is shown. The solid black points represent the lattice data \cite{Baker:2019gsi}, where the green dashed curve, the red dotted curve, and the blue dashed-dotted curve represent the results from MLP, KAN, and the Clem parameterization, respectively.}
	\label{fig:mlp-vs-kan}
\end{figure}

The comparison of both machine learning frameworks with lattice simulation results and the Clem parameterization Eq. (\ref{eq:chromofileld}) is illustrated in Fig. \ref{fig:mlp-vs-kan}, where the green dashed curve, the red dotted curve, and the blue dashed-dotted curve represent the results from MLP, KAN, and the Clem parameterization, respectively. For $d = 0.51$ fm, the parameters in Eq. (\ref{eq:chromofileld}) are set to $\varphi = 0.128,\ \alpha = 2.141,\ \mu = 6.039$ fm$^{-1}$ \cite{Baker:2019gsi}. In this work, we do not consider the uncertainties associated with the lattice simulation results and the parameterization.

The KAN training results in a target function obtained by summing multiple univariate functions, which corresponds to KART. In this testing, three univariate functions $\sin(x),\ \exp(x),\ x^2$ are derived, represented by the three red curves on the right side of Fig. \ref{fig:KAN}, listed from bottom to top. The resulting function is as follows
\begin{equation}
	\begin{aligned}
		E(d=0.51\ \mathrm{fm},x_t)=0.1744\{-0.1487\exp[1.6544\\
		\times\sin(2.6038x_t + 7.9828)]- 1\}^2 - 0.1854.
	\end{aligned}
	\label{eq:ana-kan}
\end{equation}
It can be observed that the expression above differs from Eq. (\ref{eq:chromofileld}). While Eq. (\ref{eq:chromofileld}) also converges to zero when extended beyond the range of the data on the horizontal axis, the dataset terminates at $x_t \sim 0.85$ fm. Due to the small sample size, KAN is unable to accurately predict the unknown region. If one artificially includes points that converge to zero in the training set, the training results will also exhibit a convergent function.

\subsection{Two-dimensional input training of MLP and KAN}
\label{subsec:2d}
\begin{figure*}[htbp]
	\begin{center}
		\includegraphics[width=0.98\linewidth]{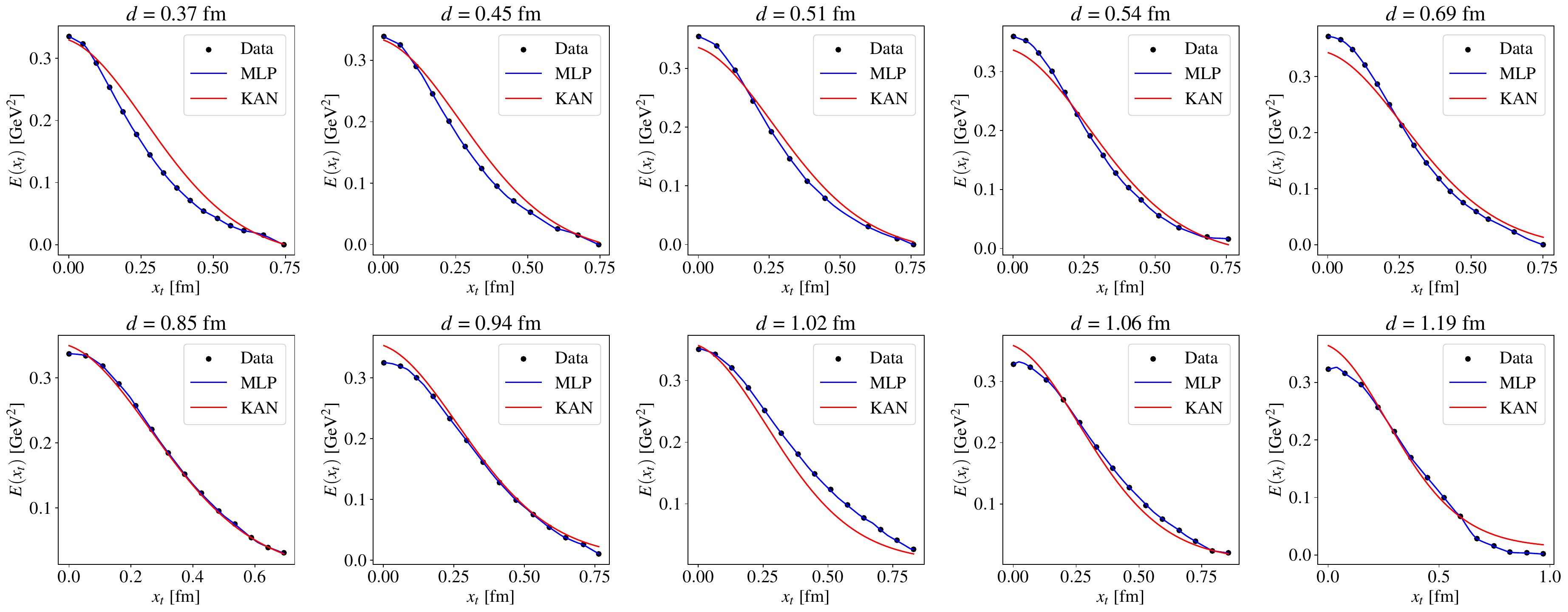}
	\end{center}
	\caption{The transverse distribution of the chromo field for quark-antiquark separation at difference $d$'s are shown. The solid black points represent the lattice data \cite{Baker:2019gsi}, where the blue and red solid curves represent the results from MLP, KAN training, respectively.}
	\label{fig:mlp-vs-kan-2d}
\end{figure*}

Now, let us focus on a more complex scenario where the training set includes the entire non-perturbative chromo field data corresponding to ten different quark-antiquark separation distances. First, we consider the MLP architecture, employing the neuron configuration shown in Fig. \ref{fig:MLP}, while maintaining the other hyperparameters consistent with the previous section. In contrast to the one-dimensional input for KAN, the complexity of the two-dimensional input increases. This is attributed to the unclear dependence of the data on $d$; the same value of $d$ can map to different $x_t$ values and their corresponding fields $E(x_t)$. Consequently, debugging the network for this training set becomes more challenging, introducing greater uncertainty. We adopt the neuron configuration scheme $(2,6,1)$, and the resulting comparison with MLP and lattice data is illustrated in Fig. \ref{fig:mlp-vs-kan-2d}.

\begin{figure}[htbp]
	\begin{center}
		\includegraphics[width=0.98\linewidth]{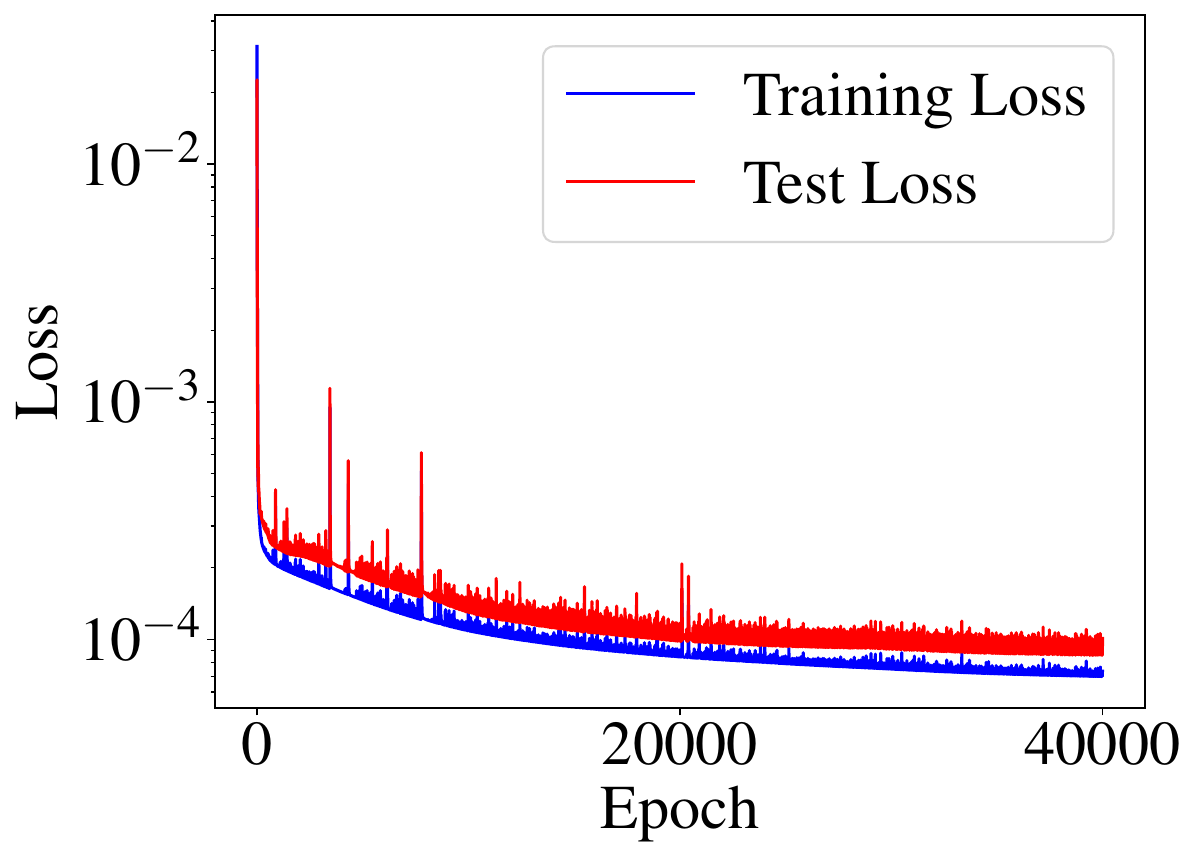}
	\end{center}
	\caption{The comparison of the loss functions between the training and test sets in the case of MLP architecture.}
	\label{fig:mlp-vs-kan-loss}
\end{figure}

Similar to Sec. \ref{subsec: 1d}, we can also obtain a symbolic expression for the KAN training results, denoted as $E(d,x_t)$. The expression corresponding to the red curve (KAN) in Fig. \ref{fig:mlp-vs-kan-2d} is as follows
\begin{equation}
	\begin{aligned}
	E(d,x_t)=0.0423d- 0.0388 + 0.0103\\
	\times\exp[-3.5478\sin(1.68x_t + 4.791)].
\end{aligned}
	\label{eq:2d ana}
\end{equation}
By comparing Eqs. (\ref{eq:ana-kan}) and (\ref{eq:2d ana}), it is evident that both of them contain the univariate functions $\sin(x)$ and $\exp(x)$; however, Eq. (\ref{eq:2d ana}) introduces another variable $d$ while omitting the quadratic function. Additionally, we observe that the training results of the KAN architecture with two-dimensional input are slightly inferior to those of the MLP architecture, as reflected in the convergence of the loss functions. Similar to the univariate MLP, the training and testloss functions for the bivariate MLP both converges to approximately $10^{-5}$ (see Fig. \ref{fig:mlp-vs-kan-loss}), while in the case of KAN, the convergence is only around $10^{-2}$. We acknowledge that this discrepancy is strongly related to the hyperparameter dependence in the KAN training process. We also experimented with different hyperparameter combinations; one alternative option involves using a $(2,4,1)$ neuron configuration, which led to results that are independent of $d$. The training set itself exhibits limited dependence on $d$, as mentioned in the corresponding lattice studies \cite{Baker:2019gsi,Baker:2024peg,Baker:2024rjq}, which is also reflected in the calculations of the flux tube string tension and the root mean square width based on the chromo field.

At this point, we have obtained the MLP and KAN results trained using lattice simulation data, making it necessary to calculate the corresponding flux tube string tension of the chromo field. In Ref. \cite{Baker:2019gsi}, the authors also provide results for the flux tube string tension, with the calculation method outlined in the square root of Eq. (\ref{eq:tension}). Additionally, to gain further insights into the flux tube structure, we calculate the root mean square width:
\begin{equation}
	\sqrt{w^2}=\sqrt{\frac{\int d^2x_t x_t^2E(x_t)}{\int d^2x_t E(x_t)}}.
	\label{eq: width}
	\end{equation}

\begin{figure*}[hbpt]
	\begin{center}
		\includegraphics[width=0.48\linewidth]{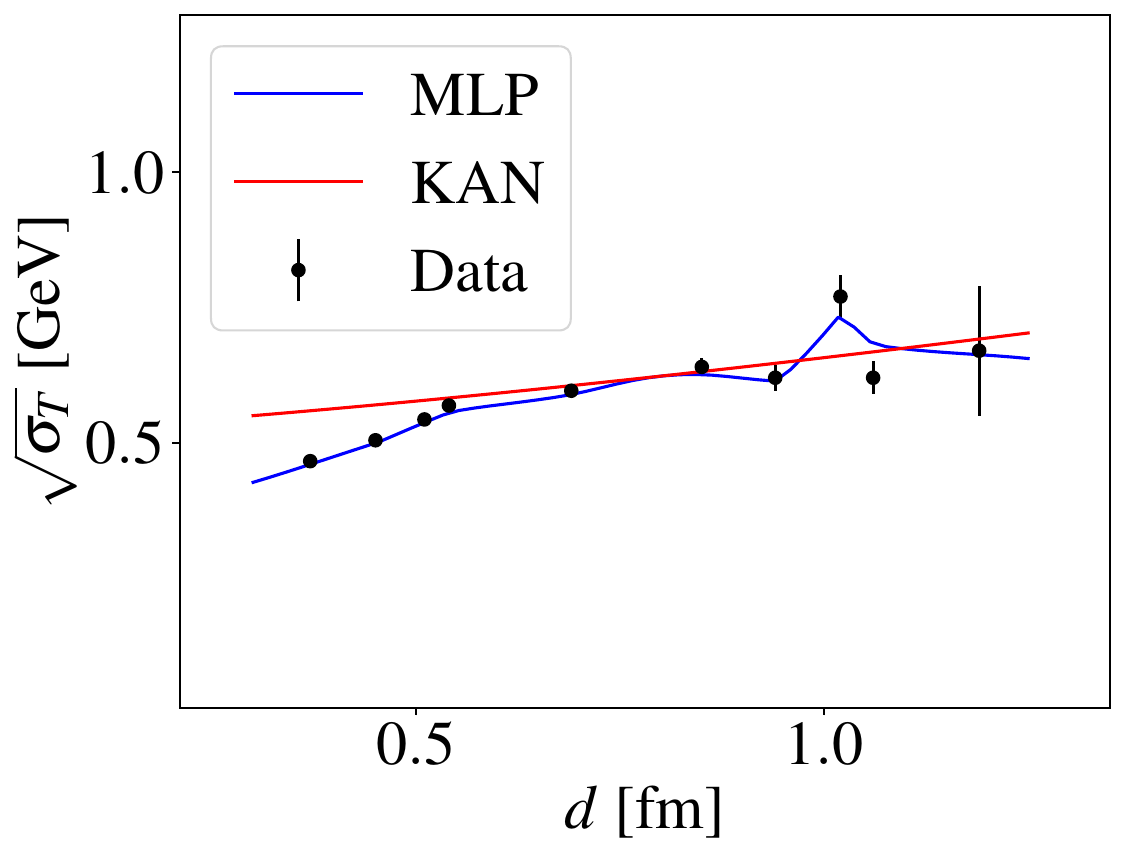}
		\includegraphics[width=0.48\linewidth]{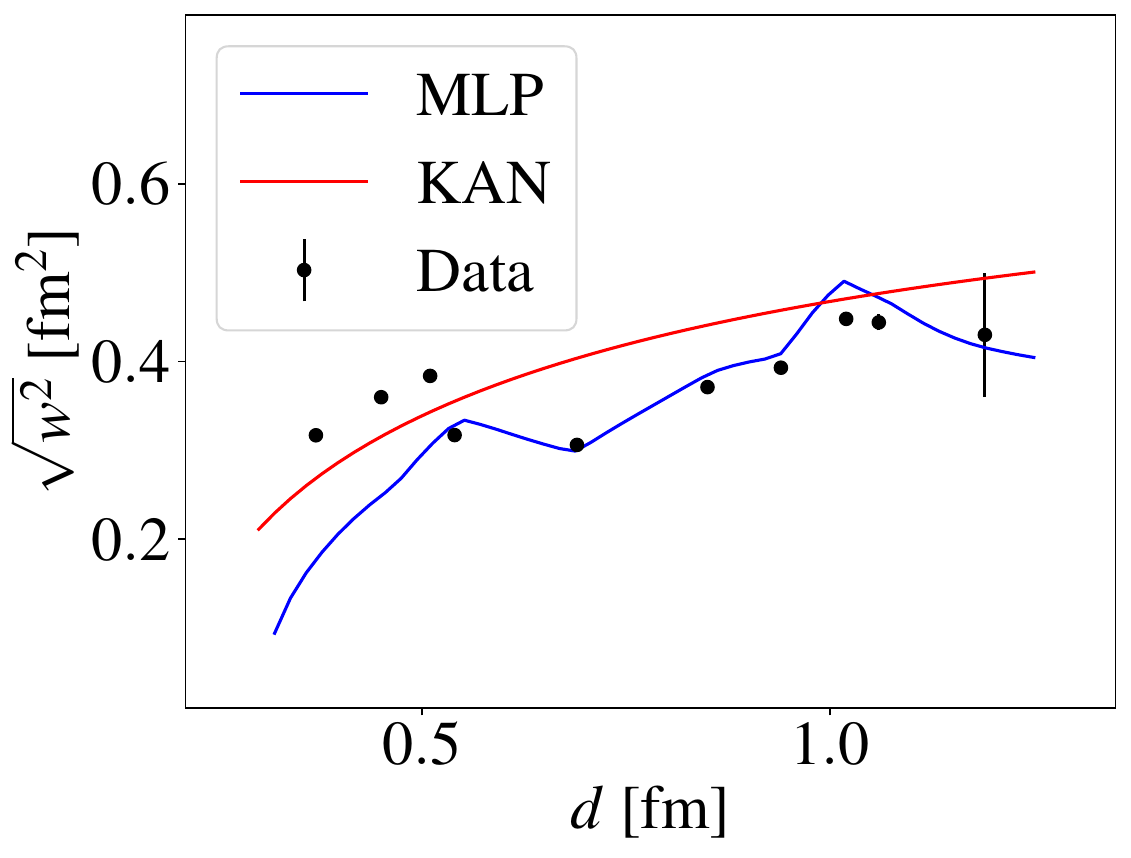}
	\end{center}
	\caption{The square root of the string tension and flux tube width obtained using MLP and KAN training, compared with lattice simulation \cite{Baker:2019gsi}. The legend is the same as that in Fig. \ref{fig:mlp-vs-kan-2d}.}
	\label{fig:mlp-vs-kan-sigma}
\end{figure*}

In Fig. \ref{fig:mlp-vs-kan-sigma}, we present the results of the chromo field calculated using both machine learning architectures to determine the flux tube string tension and flux tube width, compare them with the lattice results. Once again, the solid blue and red curves represent the numerical calculations of Eqs. (\ref{eq:tension}, \ref{eq: width}) obtained from the MLP and KAN approaches, respectively. It is important to emphasize that the root mean square width of the flux tube obtained from the MLP results is somewhat less accurate compared to that derived from the root mean square string tension, particularly in the region of small quark-antiquark separation $d$. This is primarily due to the presence of local oscillations or divergences in the higher-order moments of the chromo field generated by the MLP approach, which can be addressed by adjusting the regularization scheme during the training process.

It becomes apparent that, in the computation of string tension, MLP continues to outperform the KAN architecture. This is not due to an inherent flaw in KAN but rather stems from the uncertainties introduced in tuning the hyperparameters of KAN. We emphasize that the impact of hyperparameter adjustments on machine learning outcomes is observable; however, the study of hyperparameters themselves is beyond the scope of this work. Determining the hyperparameters for neural network architectures is a substantial undertaking, especially for the KAN architecture, where optimizing these parameters is crucial for its applications in physics. Currently, there is a popular and open-source hyperparameter optimization library known as Keras Tuner \cite{omalley2019kerastuner}, which includes built-in algorithms for Bayesian Optimization, Hyperband, and Random Search. Recent studies \cite{Almaeen:2022imx,Almaeen:2024guo} have utilized this approach to optimize their own neural network architectures, and interfacing Keras Tuner with KAN may address the complexities encountered in tuning KAN's hyperparameters, which is an endeavor we plan to pursue in the future.

\section{Summary and outlook}
\label{sec:conclusion}
Understanding the spatial structure of gauge fields, particularly the flux tube model, is essential for elucidating the interactions between quark-antiquark pairs and the associated phenomena of quark confinement, including string breaking. In this study, we leverage machine learning techniques, notably MLP and the KAN, to analyze flux tube structures simulated via lattice QCD. Our objective is to harness AI to reproduce the lattice results effectively. We derive analytical expressions that characterize the chromo field distributions for varying quark-antiquark separations using the KAN method. We find that the MLP architecture adeptly captures the salient features of lattice data, while the KAN strategy involves greater complexity, particularly in replicating two-dimensional training datasets. Nonetheless, we successfully establish a two-dimensional expression for the chromo field distribution that closely mirrors lattice results. Previously, the parameterization of chromo fields based on the second type of superconducting duality could only provide a set of parameters corresponding to a fixed quark-antiquark separation $d$. In other words, while these models describe lattice results well, they are limited in predicting the chromo field profiles for unknown values of $d$. We believe that applying a neural network deep learning approach to analyze the existing limited lattice simulations can contribute positively to predicting the chromo field distributions for different values of $d$. 

Additionally, we compute the variations of the flux tube string tension as a function of quark-antiquark separation, comparing outcomes from both trained MLP and KAN models. Our results reveal that the MLP approach outperforms KAN in operational efficiency and adaptability to large-scale datasets, while KAN offers advantages in generating interpretable analytical expressions with fewer assumptions. Despite the necessity for extensive optimization of KAN's hyperparameters, we suggest the integration of adaptive hyperparameter tuning interfaces, such as Keras Tuner, to enhance the model's applicability for learning physical phenomena. This work exemplifies the potential of AI methodologies in high-energy physics, contributing to the understanding of fundamental interactions via advanced computational techniques.

\section*{Acknowledgments}
We greatly appreciate the discussions with Dr. Xun Chen. This work has been supported by the Strategic Priority Research Program of Chinese Academy of Sciences (Grant NO. XDB34030301) and National Key R$\&$D Program of China (Grant NO. 2024YFE0109800 and 2024YFE0109802).

%\goodbreak

\bibliography{refs}

\end{document}